\documentclass[epj]{svjour}
\usepackage{latexsym}
\usepackage{graphics}
\usepackage{amsmath}
\usepackage{dcolumn}
\newcommand{\be}{\begin{equation}}
\newcommand{\ee}{\end{equation}}
\newcommand{\ba}{\begin{eqnarray}}
\newcommand{\ea}{\end{eqnarray}}
\newcommand{\sba}{\begin{subequations}}
\newcommand{\sea}{\end{subequations}}
\newcommand{\barr}{\begin{array}}
\newcommand{\earr}{\end{array}}
\newcommand{\nn}{\nonumber \\}
\newcommand{\bm}{\begin{mathletters}} 
\newcommand{\eml}{\end{mathletters}}

\newcommand{\dl}{\delta}

\def\C {{{\cal C}}}
\def\D {{{\cal D}}}

\def\H {{{\cal H}}}

\def\N {{{\cal N}}}

\def\X {{{\cal X}}}
\def\Y {{{\cal Y}}}

\begin{document}

\title{Mean-Field Theory for Fermion Pairs and the ab initio Particle-Vibration-Coupling Approach }
\author{Peter Schuck\inst{1} %\and Second author\inst{2}% etc
\thanks{\emph{Present address:} Institut de Physique Nucl\'eaire d'Orsay, Universit\'e Paris-Sud, CNRS--IN2P3. 15, Rue Georges Clemenceau, 91406 Orsay Cedex, France.}%
}                     % Do not remove
\institute{Institut de Physique Nucl\'eaire Orsay \and Univ. Grenoble Alpes, CNRS, LPMMC, 38000 Grenoble, France}
\date{Received: date / Revised version: date}
\abstract{
A Dyson Bethe-Salpeter equation (Dyson-BSE) for fermion pairs is presented whose kernel has a static and a one frequency dependent contribution, analogous to the self energy of the single particle Dyson equation with the (static) mean field term and the energy dependent correlation term. The static part of the Dyson-BSE is the self-consistent mean field for the vibrations. At the same time, for the correlated single particle self-energy a full particle-vibration coupling (PVC) scattering equation is established where the vibration is the same as obtained from the Dyson-BSE. Both equations, single particle Dyson equation and Dyson-BSE, are coupled through self-consistency. Numerical results for Lipkin and 1D Hubbard chain are very promising.
\PACS{
      {PACS-key}{discribing text of that key}   \and
      {PACS-key}{discribing text of that key}
     } % end of PACS codes
} %end of abstract
\maketitle

\section{Introduction}

Nuclei are a paradigmatical case where single nucleon properties, response functions, and pair propagators are strongly influenced by collective modes. In order to cope with this situation, the so-called particle-vibration coupling (PVC) approach has been put forward by P. F. Bortignon and collaborators \cite{Colo}. The PVC approach is based on effective nucleon forces of the Skyrme or Gogny type in the non-relativstic scheme or on effective density dependent coupling constants in the relativistic theory. In this sense PVC is still largely phenomenological.\\
The objective of this work is to present a PVC approach which can be used in ab initio calculations where one starts from a given bare force. Such systems are, e.g., electron systems with the Coulomb force and the coupling of single particle (s.p.) motion to the plasmon resonance but one can also think to start for nuclear systems with a 'bare' force which reproduces very well nucleon-nucleon phase shifts as does for example the Bonn force \cite{CD-Bonn}. In order to achieve this, we will make use of a mean field approach for correlated fermion pair operators of the type $(a^+a)$ or $(aa)$ and $(a^+a^+)$ with $a^+,a$ fermion creation and destruction operators. We will show that the corresponding correlated ground state wave function is given to good approximation by the so-called Coupled-Cluster-Doubles (CCD) wave function \cite{jemai13}. The collective modes obtained in this way can be used to set up a particle-vibration scattering equation entering the self-energy of the single fermion Dyson equation, so that via the s.p. occuaption numbers the single particle and two particle propagators get selfconsistently coupled. The performance of this approach will be demonstrated with model cases where a comparison with exact solutions is possible.

\section{ The formalism}

The mean-field approach for fermion pairs is, in principle, not new and we will recapitulate it here only very succinctly. We will present it, establishing a formally exact Bethe-Salpeter equation (BSE) whith a one frequency kernel which has a definite expression. The point we want to make here is that in others than the nuclear field a one frequency BSE is very uncommon and mostly propagators with four times (or three times after translational invariance in time is taken into account) are considered implying that also the integral kernel depends on three frequencies. However, since this leads to numerically inextractable equations, mostly the three time (or three frequency) kernel is taylored down either to a static one or to a one frequency kernel making use of some suitable approximations. We have recently published a longer article where this problematic is discussed at length in the particle-hole (ph) channel in the context of electronic systems \cite{Julien}. In this work we will discuss, additionally, the particle-particle (pp) channel with its possible pairing instability. However, we also will expose specific aspects in the ph channel and in any case pp and ph channels are coupled, e.g. via screening of the bare interaction in both channels. As mentioned in the Introduction, a novelty will also be that we set up a single particle self-energy which is consistent with the mean field approach of fermion pairs. Notably a particle-vibration scattering equation will be established.\\
Let us start with what we want to call a formerly exact Dyson-Bethe-Salpeter equation (Dyson-BSE) for the two time pair propagator

\begin{equation}
  G^{t-t'}_{k_1k_2k'_1k'_2} = -i\langle 0|{\rm T}(c_{k_1}c_{k_2})_t(c^+_{k'_2}c^+_{k'_1})_{t'}|0\rangle
  \label{G-2}
\end{equation}
Here T is the time ordering operator, and $|0\rangle $ stands for the exact ground state and the fermion pair operators turn in time with the two body Hamiltoinan

\begin{eqnarray}
  &H&= H_0 + V \equiv \nonumber\\
  \sum_{kk'}&e_{kk'}&c^+_kc_{k'} + \frac{1}{4}\sum_{k_1k_2k_3k_4}
\bar v_{k_1k_2k_3k_4}c^+_{k_1}c^+_{k_2}c_{k_4}c_{k_3}
\label{H}
\end{eqnarray}
where $e_{kk'}$ is the single particle matrix comprising kinetic energy and external potential, and

\begin{equation}
  \bar v_{k_1k_2k_3k_4} = \langle k_1k_2|v|k_3k_4\rangle - \langle k_1k_2|v|k_4k_3\rangle
  \label{as-v}
  \end{equation}
is the antisymmetrized matrix element of the two body interaction. \\
The two-body propagator obeys the following exact Dyson-BSE \cite{Julien,NPA628}

\begin{eqnarray}
  &(& i \partial_t -\tilde e_{k_1} - \tilde e_{k_2})G^{t-t'}_{k_1k_2,k'_1k'_2} = N_{k_1k_2k'_1k'_2}\delta(t-t')\nonumber\\
  &+& \sum_{k_3k_4}\int dt_1 [K^{pp,0}\delta(t-t_1) + K^{pp,{\rm dyn.},t-t_1}]_{k_1k_2k_3k_4}\nonumber\\
  &&N^{pp-1}_{k_3k_4}G^{t_1-t'}_{k_3k_4k'_1k'_2}
\label{ppBSE}
\end{eqnarray}
where with $\delta_{k_1k_2,k'_1k'_2} =\delta_{k_1k'_1}\delta_{k_2k'_2} - \delta_{k_1k'_2}\delta_{k_2k'_1}$
\begin{equation}
  N^{pp}_{k_1k_2k'_1k'_2} = \delta_{k_1k_2,k'_1k'_2}N^{pp}_{k_1k_2};~~~N^{pp}_{k_1k_2} =1-n_{k_1}-n_{k_2}
  \label{pp-occ}
\end{equation}
%and
%\begin{equation}
%  N^{pp}_{k_1k_2} =1-n_{k_1}-n_{k_2}
%  \label{pp-occ}
%\end{equation}
and we supposed that we work in the canonical basis where the density matrix is diagonal, that is

\[ \langle 0| c^+_{k_1}c_{k'_1}|0\rangle = \delta_{k_1k'_1}n_{k_1} \]
Furthermore the s.p. energies in (\ref{ppBSE}) are given by
  
\begin{equation}
  \tilde e_k = e_k + V^{\rm MF}_k
  \label{HF-e}
\end{equation}
where the mean field shift is included

\begin{equation}
  V_k^{{\rm MF}} = \langle 0|\{c_k,[H,c^+_k]\}|0\rangle = \sum_{k'}\bar v_{kk'kk'}n_{k'}
  \label{MF}
  \end{equation}
and where $\{..\}$ stands for the anticommutator. We assumed that mean-field energies and density matrix can be diagonalized simultaneously.
  The integral kernel is given by

\begin{eqnarray}
 K^{pp}_{k_1k_2k'_1k'_2} &=& \langle [A_{k_1k_2},[V,A^+_{k'_1k'_2}]]\rangle\delta(t-t')\nonumber\\
&+& (-i)\langle {\rm T}  J_{k_1k_2}(t) J^+_{k'_1k'_2}(t')\rangle_{\rm irr.}\nonumber\\
&\equiv& K^{pp,0}_{k_1k_2k'_1k'_2}\delta(t-t') +  K^{pp,{\rm dyn.}, t-t'}_{k_1k_2k'_1k'_2}
\label{kernel2}
\end{eqnarray}
where we abreviated

\begin{equation}
  A_{k_1k_2} = c_{k_1}c_{k_2}
\end{equation}
and

\begin{equation}
  J^{pp}_{k_1k_2}=[A_{k_1k_2}, V]~~ = j_{k_1}c_{k_2} +c_{k_1}j_{k_2}
  \label{pp-current}
\end{equation}
with
\begin{equation}
j_k = 
[c_k,V] = \frac{1}{2}\sum_{k_2k_3k_4}\bar v_{kk_2k_3k_4}c^+_{k_2}c_{k_4}c_{k_3}\,.
\label{j-k}
  \end{equation}

Please note that the $K$ matrix which after Fourier transform depends only on one frequency can be interpreted as a self-energy for the motion of a fermion pair. As usual the self energy is split into a frequency independent, static part and a truely frequency dependent, dynamic part. The latter must be two line irreducible, hence the index  'irr.'. This partition is in complete analogy to the case of the s.p. self-energy appearing in the s.p. Dyson equation for the s.p. Green's function
\begin{equation}
G^{t-t'}_{kk'} = -i\langle 0|{\rm T}c_k(t)c^+_{k'}(t')|0\rangle
\label{GF1}
\end{equation}
The Dyson equation then reads

\begin{equation}
(i\partial_t - e_k) G^{t-t'}_k = \delta(t-t') +\int dt_1 \Sigma_k^{t-t_1}G^{t_1-t'}_k
\label{Dyson}
\end{equation}
with the self-energy expressed as

\begin{equation}
 \Sigma_k^{t-t'} = V_k^{{\rm MF}}\delta(t-t') -i\langle 0|{\rm T}j_k(t)j_k^+(t')
|0\rangle_{{\rm irr.}} 
\label{self4}
\end{equation}
where the index 'irr.' again indicates that the corresponding correlation function should be one-line irreducible.

%\begin{equation}
 % n_k = \langle 0|c^+_kc_k|0\rangle
 % \label{occs}
 % \end{equation}
%are the occupation numbers.
It is well known that the self-energy of the s.p. motion is closely related to the optical potential of elastic nucleon-nucleon scattering \cite{Bell-Squ}. Now suppose one wants to consider elastic deuteron-nucleus scattering (or any elastic scattering of composite bosons) how would one define a corresponding optical potential which by definition will depend only on the incoming deuteron energy ? Of course quite naturally the deuteron optical potential can directly be derived from the one frequency $K$-matrix of the Dyson-BSE defined above, see an early work on this in Ref. \cite{opt-pot}.

\subsection{Static part of the BSE kernel}

Let us now discuss the $K^{pp,0}$ term of the BSE kernel. To establish an explicit form for $K^{pp.0}$, we have to evaluate the double commutator contained in the pair mean field part of $ K^{pp,0}$ \cite{NPA628}

\begin{eqnarray}
  K^{pp,0}_{k_1k_2k'_1k'_2} &=& N^{pp}_{k_1k_2}\bar v_{k_1k_2k'_1k'_2}N^{pp}_{k'_1k'_2}
  \nonumber\\
  &&+\biggl \{\bigg [ \bigg (\frac{1}{2}\delta_{k_1k'_1}\bar v_{l_1k_2l_3l_4}C_{l_3l_4k'_2l1}
  \nonumber\\
    &&+\bar v_{l_1k_2l_4k'_2}C_{l_4k_1l_1k'_1}\bigg )-(k_1 \leftrightarrow k_2)\bigg ] 
    \nn
     &&~~~~~ - (k'_1 \leftrightarrow k'_2]\biggr \}
      \label{Kpp-0}
\end{eqnarray}
where

\begin{equation}
  C_{k_1k_2k'_1k'_2} = \langle 0|c^+_{k'_1}c^+_{k'_2}c_{k_2}c_{k_1}|0\rangle -n_{k_1}n_{k_2}
    \delta_{k_1k_2k'_1k'_2}
    \label{C-2}
\end{equation}
which is the fully correlated two body or cumulant form of the density matrix.\\

We see that $K^{pp,0}$ involves, besides occupation numbers, static two-body correlation functions. They are of two types: there are single line corrections with one of the two s.p. motions uneffected by the correlations (those with the Kronecker symbols) and there are exchange terms where a two-body correlation is exchanged between the two particles. Since our starting point is a two-body propagator, a self-consistent scheme can be established. This is similar to the selfconsistency involved with  the s.p. mean-field, only here, naturally, two-body correlation functions have to be iterated rather than s.p. densities in the case of the s.p. mean-field. We, therefore, call $K^{pp,0}$ the 'fermion pair mean field'. Of course, there appear also s.p. densities in $K^{pp,0}$ and we will later show how they can be consistently obtained from the s.p. Green's function.\\
A closer investigation of the exchange kernel, however, shows that the exchange is rather of the ph type. At least to lowest order, that is to second order, the exchange is given by a static ph exchange. This ph exchange is well known that it screens the pairing force by almost a factor of two as has first been evaluated by Gorkov, Melik-Barkhudarov (GMB) \cite{GMB,Strinati}. Actually for systems like the nuclear ones where there are more than two species of fermions (that is, four), the screening can also become anti-screening \cite{Pethick,Urban}.
Also GMB did not use strict second order but replaced the vertices by the scattering length, that is the vertices have been dressed to $T$-matrices in the low energy limit, that is, by the scattering length. Though this resummation can be understood easily by graphical analysis, how this can be derived more analytically will be discussed below in sect. II.B. \\
The first term on the r.h.s. of (\ref{Kpp-0}) is the usual two body matrix element of the interaction modified with correlated occupation numbers ( the standard particle-particle RPA as described in \cite{RS} uses HF occupation numbers). One can presum this term what leads to the so-called renormalized pp-RPA \cite{Hirsch}

\begin{eqnarray}
  G^{{\rm r-ppRPA}}_{k_1k_2 k'_1k'_2} = G^{0,{\rm r-ppRPA}}_{k_1k_2k'_1k'_2} +\sum_{k_3k_4} G^{0,{\rm r-ppRPA}}_{k_1k_2k_3k_4}\bar v_{k_3k_4k'_3k'_4}
  \nn
  .G^{{\rm r-ppRPA}}_{k_3k_4 k'_1k'_2}
  \label{r-ppRPA}
\end{eqnarray}
with

\begin{equation}
  G^{0,{\rm r-ppRPA}}_{k_1k_2k'_1k'_2} = \frac{1 -n_{k_1}-n_{k_2}}{\omega - e_{k_1}-e_{k_2}}
  \delta_{k_1k_2k'_1k'_2}
 \label{0-r-ppRPA}
\end{equation}
The Dyson-BSE can then be written in the following way

\begin{eqnarray}
  G_{k_1k_2k'_1k'_2} &=& G^{{\rm r-ppRPA}}_{k_1k_2 k'_1k'_2} 
  \nn 
  &&+ G^{{\rm r-ppRPA}}_{k_1k_2 k_3k_4}[ N^{-1}(K^{pp,0} - N\bar vN
\nn
    &&+ K^{{\rm dyn.}})N^{-1}]_{k_3k_4k'_3k'_4}G_{k'_3k'_4k'_1k'_2}
  \label{dressed-BSE}
\end{eqnarray}

\subsection{Dynamic part of the BSE kernel}

\begin{figure}
\resizebox{0.4\textwidth}{!}{\includegraphics{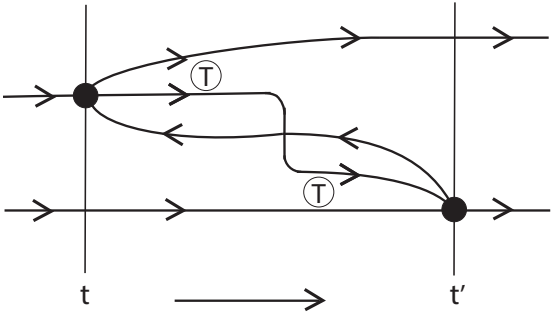}}
  \caption{ Second order $T$-matrix approximation to the three particle propagator contained in the 3p-1h correlation function. Together with the first order contribution contained in the uncorrelated 3p-1h propagator, this can be summed to one full $T$-matrix. Time flows from left (t) to right (t').}
  \label{GMB-T}
\end{figure}

Let us now discuss the time-dependent, {\it  dynamic} part $K^{\rm dyn.}$ of the interaction kernel 

\begin{equation}
  K^{pp,\rm dyn.}_{k_1k_2k'_1k'_2} = -i \langle 0|TJ_{k_1k_2}(t)J^+_{k'_1k'_2}(t')|0\rangle_{\rm irr.}
  \label{K-dyn}
\end{equation}
From (\ref{pp-current}) we see that this expression involves four different contributions: two contributions contain the two interaction vertices on the same line and two contributions on opposite lines. The latter, therefore, contain exchange processes while the former are responsible for s.p. self energy corrections. Approximating the the 3p-1h propagator involved in (\ref{K-dyn}) by a product of a hole propagator and the three body propagator in second order $T$-matrix approximation, we give a schematic grapphical representation of the term in
Fig.\ref{GMB-T}. This illustrates how one can replace in the second order screening term discussed above, the bare vertices by ladder $T$-matrices and then eventually by the scattering lengths as done in \cite{GMB}. We see that the exchange contributions are of the screening (or anti-screening) type whereas the other two contributions renormalize the s.p. by particle-vibration couplings. Of course, in general, all four lines are correlated.\\
As a matter of fact the 3p-1h propagator in (\ref{K-dyn}) lends itself to several ``natural'' approximations other than the one we just discussed. For example instead of considering an uncorrelated ph-propagator exchange, one could take the ph-response function which will be presented in sect. II.C \cite{Lombardo,Urban}.

\subsection{The ph-channel}

Since we see that ph and pp channels are coupled, we immediately also give the Dyson-BSE in the ph-channel \cite{Julien}

\begin{eqnarray}
  &&
  ( \omega -\tilde \epsilon_{k_1} + \tilde \epsilon_{k_2})R_{k_1k_2k'_1k'_2}(\omega) = N^{ph}_{k_1k_2k'_1k'_2}
  \nonumber \\ &&
  + \sum_{k_3k_4} [K^{ph,0}_{k_1k_2k_3k_4}  + K^{ph,\mathrm{dyn}}_{k_1k_2k_3k_4} (\omega)]N^{-1}_{k_3k_4} R_{k_3k_4k'_1k'_2}(\omega).
\label{BSE}
\end{eqnarray}  
With

\begin{equation}
  N^{ph}_{k_1k_2k'_1k'_2} = \delta_{k_1,k_2}N^{ph}_{k_1k_2}\equiv \delta_{k_1,k_2}(n_{k_2}-n_{k_1})
\end{equation}
and 
the two-time response function defined by (with $k_1 \ne k_2$ and $k_1' \ne k_2'$)
\begin{equation}
 R_{k_1k_2k'_1k'_2}(t-t') =  -i \langle 0| \mathrm{T}\{ c^\dag_{k_2}(t)c_{k_1}(t)
c^\dag_{k'_1}(t')c_{k_2'}(t') \} |0\rangle 
\label{response}
\end{equation}
The static part of the integral kernel is given by
\begin{eqnarray}
  &&
    K^{ph,0}_{k_1k_2k_3k_4} = N^{ph}_{k_1k_2}\bar v_{k_1k_4k_2k_3}N^{ph}_{k_3k_4}
  \nonumber\\ && \quad
  \Big[- \frac{1}{2}\sum_{ll'l''}(\delta_{k_2k_4} \bar v_{k_1ll'l''}C_{l'l''k_3l} +\delta_{k_1k_3}\bar v_{ll'k_2l''}C_{k_4l''ll'})
  \nonumber\\ && \quad
  + \sum_{ll'}(\bar v_{k_1lk_3l'}C_{k_4l'k_2l} + \bar v_{k_4lk_2l'}C_{k_1l'k_3l})
  \nonumber\\ && \quad
  - \frac{1}{2}\sum_{ll'}(\bar v_{k_1k_4ll'}C_{ll'k_2k_3} + \bar v_{ll'k_2k_3}C_{k_1k_4ll'})\Big],
\label{K-ph0}
\end{eqnarray}
and the dynamic part

\begin{eqnarray}
    K^{ph,\rm dyn}_{k_1k_2k'_1k'_2}(t-t')&=&
    -i\langle 0 | \mathrm{T} \{  J^{ph}_{k_1k_2}(t)  J^{ph\dag}_{k'_1k'_2}(t')\} | 0 \rangle^\mathrm{irr}.\nonumber\\
\label{kernel2}
\end{eqnarray}
with

\begin{eqnarray}
  J^{ph}_{k_1k_2} &=&[c^{\dag}_{k_2}c_{k_1},V]
 \nonumber\\
  &=&  c^\dag_{k_2}j_{k_1} + j^\dagger_{k_2}c_{k_1}
    ,
  \label{J-kk'}
\end{eqnarray}
As in the pp-channel, we can introduce a renormalized ph-propagator
\begin{equation}
  R^{{\rm r-phRPA}}_{k_1k_2 k'_1k'_2} = R^{0,{\rm r}}_{k_1k_2k'_1k'_2} +\sum_{k_3k_4} R^{0,{\rm r}}_{k_1k_2k_3k_4} T_{k_3k'_4k_4k'_3}R^{{\rm r-phRPA}}_{k'_3k'_4 k'_1k'_2}
  \label{r-ppRPA}
\end{equation}
with

\begin{equation}
  R^{0,{\rm r}}_{k_1k_2k'_1k'_2} = \frac{n_{k_1}-n_{k_2}}{\omega - e_{k_1}+e_{k_2}}
  \delta_{k_1k'_1}\delta_{k_2k'_2}
 \label{0-r-ppRPA}
\end{equation}
and where we introduced the ladder $T$-matrix

  \begin{eqnarray}
    T_{k_1k_2k_3k_4} &=& \bar v_{k_1k_4k_2k_3}
  \nonumber\\ && 
  N^{-1}_{k_1k_2}\Big[
    - \frac{1}{2}\sum_{ll'}(\bar v_{k_1k_4ll'}C_{ll'k_2k_3}
\nn
    &+& \bar v_{ll'k_2k_3}C_{k_1k_4ll'})\Big]N^{-1}_{k_3k_4},
\label{K-0}
\end{eqnarray}
which is important when dealing with systems with a hard-core potential.\\

\begin{figure}
\resizebox{0.45\textwidth}{!}{\includegraphics{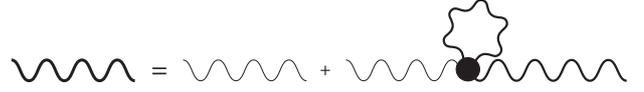}}
  \caption{Mean-field ph-propagator with tad-pole self-interaction}
  \label{Bo-tadpole}
  \end{figure}

Keeping from the ph-kernel only the remaining instantaneous part, one arrives at a self-consistent mean-field equation for the ph-propagation

\begin{eqnarray}
  R^{{\rm MF-RPA}}_{k_1k_2 k'_1k'_2} &=& R^{{\rm r-phRPA}}_{k_1k_2k'_1k'_2}
\nn
&+&\sum_{k_3k_4k'_3k'_4} R^{{\rm r-phRPA}}_{k_1k_2k_3k_4}[N^{-1}\tilde K^{ph,0}N^{-1} ]_{k_3k_4k'_3k'_4}
\nn
&& ~~~~R^{{\rm MF-RPA}}_{k'_3k'_4 k'_1k'_2}
  \label{mf-phRPA}
\end{eqnarray}
where $\tilde K^{ph,0}$ is the part of (\ref{K-ph0}) where only the second, third, fourth, and fifth terms are kept.\\
The eigenvalue form of this equation, given below, is known as the self-consistent RPA (SCRPA) equation \cite{jemai13,NPA628}. A graphical representation of Eq.(\ref{mf-phRPA}) is given in Fig.\ref{Bo-tadpole}

We now want to give the spetral representation of the response function, since it may be important for the following when we discuss the particle-vibration coupling (PVC). 

\begin{eqnarray}
  R_{k_1k_2k'_1k'_2}(\omega) &=& \sum_{\nu}\frac{\begin{pmatrix}X^{\nu}\\Y^{\nu}\end{pmatrix}_{k_1k_2}(X^{\nu \dag}~~Y^{\nu ~~\dag})_{k'_1k'_2}}{\omega - \Omega_{\nu} + i\eta}\nonumber\\
  &-&\frac{\bigg [ \begin{pmatrix}Y^{\nu}\\X^{\nu}\end{pmatrix}_{k_1k_2}(Y^{\nu \dag} ~~X^{\nu \dag})_{k'_1k'_2}\bigg ]^*}{\omega + \Omega_{\nu} - i\eta}
  \label{spectral}
  \end{eqnarray}
On the pole we get with (\ref{spectral}) the following eigenvalue equation (SCRPA equation \cite{jemai13})

%\begin{equation}
%  {\mathcal S}\chi = \Omega \chi
%  \label{short-rpa}
%\end{equation}
%or more explicitly

\begin{equation}
  \begin{pmatrix}A&B\\-B^*&-A^* \end{pmatrix}
  \begin{pmatrix}X\\Y \end{pmatrix} = \Omega \begin{pmatrix}X\\Y \end{pmatrix}
  \label{RPA-eq}
\end{equation}
where

\begin{eqnarray}
  A_{k_1k_2,k'_1k'_2} &=& \frac{\langle 0|\{c^{\dag}_{k_2}c_{k_1},[H,c^{\dag}_{k''_1}c_{k'_2}]\}|0\rangle}{ \sqrt{(n_{k_2}-n_{k_2})(n_{k'_2}-n_{k'_1})}}\nonumber\\
  B_{k_1k_2,k'_1k'_2} &=& -\frac{\langle 0|\{c^{\dag}_{k_2}c_{k_1},[H,c^{\dag}_{k'_2}c_{k'_1}]\}|0\rangle}{ \sqrt{(n_{k_2}-n_{k_1})(n_{k'_2}-n_{k'_1})}}
  \label{A,B}
\end{eqnarray}
Please let us note that the same equations can be derived from a different perspective. Considering the average excitation energy given by the normalised energy weighted sum rule

\begin{equation}
  \Omega =\frac{1}{2} \frac{\langle 0|[Q,[H,Q^{\dag}]]|0\rangle}{\langle 0|[Q,Q^{\dag}]|0\rangle}
  \label{sum-rule}
\end{equation}
with

\begin{equation}
  Q^{\dag}_{\mu} = \sum_{k_1 \textgreater k_2}[X^{\mu}_{k_1k_2}\frac{c^{\dag}_{k_1}c_{k_2}}{\sqrt{n_{k_2}-n_{k_1}}} - Y^{\mu}_{k_1k_2}\frac{c^{\dag}_{k_2}c_{k_1}}{\sqrt{n_{k_2}-n_{k_1}}}]
  \label{Q}
\end{equation}
and its inversion

\begin{equation}
c^{\dag}_{k_1}c_{k_2} = \sqrt{n_{k_2}-n_{k_1}}  \sum_{\mu}[X^{\mu}_{k_1k_2}Q^{\dag}_{\mu} + Y^{\mu}_{k_1k_2}Q_{\mu}]  
\label{inversion1}
\end{equation}
With the excitation creation operator (\ref{Q}) we suppose that the excited state is given by $Q^{\dag}_{\mu}|0\rangle = |\mu\rangle$ and that there exists the so-called killing condition

\begin{equation}
  Q_{\mu}|0\rangle = 0
  \label{killing}
  \end{equation}

Minimizing the sum rule with respect to the $X,Y$ amplitudes, again leads to the eigenvalue problem (\ref{RPA-eq}).

Evaluating the double commutators with the help of (\ref{inversion1},\ref{killing}), we find that $A,B = A,B[X,Y,n_k]$, that is, the SCRPA matrix is a functional of the RPA amplitudes $X, Y$ and the occupation probabilities $n_k$ where we made the approximation that $\langle \hat n_k \hat n_{k'}\rangle \simeq n_kn_{k'}$ what usually is a very good approximation. This approximation is not necessary but avoiding it complicates the formalism quite a bit.
We again can see that the ph-channel is coupled via the exchange to the pp-channel.\\
%As in the pp-channel, we also can resum in the ph-channel the first term on the r.h.s. of (\ref{K-0})

%\begin{eqnarray}
%&& R_{k_1k_2k'_1k'_2} = R^{0,r-phRPA}_{k_1k_2k'_1k'_2} 
%  \\
% && + \sum_{k_3k_4}R^{0,r-phRPA}_{k_1k_2k_3k_4}
%  [K^{ph,0} - N\bar vN +K^{ph, {\rm dyn.}}]_{k_3k_4k'_3k'_4}
%  \nn
%  &&~~~~.R_{k'_3k'_4k'_1k'_2}\nonumber 
%  \label{resum-v}
%\end{eqnarray}
%where

%\begin{eqnarray}
% R^{0,r-phRPA}_{k_1k_2k'_1k'_2} &=& R^{0}_{k_1k_2k'_1k'_2} 
% \\
% &&+ \sum_{k_3k_4k'_3k'_4}R^{0}_{k_1k_2k_3k_4}\bar v_{k_3k_4k'_3k'_4}R^{0,r-phRPA}_{k_1k_2k'_1k'_2}\nonumber 
% \label{r-phRPA}
%\end{eqnarray}
%and

%\begin{equation}
%  R^{0}_{k_1k_2k'_1k'_2} = \frac{n_{k_2}-n_{k_1}}{\omega - e_{k_1} -e_{k_2}}\delta_{k_1k'_1}\delta_{k_2k'_2}
%  \label{free-ph}
%  \end{equation}

Further details of the procedure and approximation schemes are presented in \cite{Julien,jemai13}. For the moment let us come to the important question how to incorporate a consistent evaluation of the s.p. occupation numbers $n_k$, a question which has remained open so far.

\section{Determination of the Single Particle Occupation Numbers}

The single particle occupation numbers being a s.p. quantity, it is natural to determine them via the s.p. propagator (\ref{GF1}). The point will be to find an approximation of the self energy which is consistent with the ground state implicitly used in our two fermion mean field equations to which we want to restrict ourselves from now on. As was already remarked earlier, the ground state corresponding to the fermion pair mean field equations is to good approximation given by the CCD ground state wave function

\begin{equation}
  |Z\rangle = e^{\frac{1}{4}\sum_{p_1p_2h_1h_2}z_{p_1p_2h_1h_2}c^{\dag}_{p_1}c^{\dag}_{p_2}c_{h_2}c_{h_1}}|{\rm HF}\rangle
  \label{Z}
\end{equation}
As has been shown in \cite{jemai13}, this state is to good approximation the vacuum to the creation operators where this time not all indices are allowed but only those lying in the ph configuration space.

\begin{equation}
  Q^{\dag}_{\mu} = \sum_{ph}[X^{\mu}_{ph}\frac{c^{\dag}_pc_h}{\sqrt{n_h-n_p}} - Y^{\mu}_{ph}\frac{c^{\dag}_hc_p}{\sqrt{n_h-n_p}}]
  \label{Q-ph}
\end{equation}
and the corresponding inversion (\ref{inversion1}).

These matters have amply been discussed in earlier literature where also the relation between the $z$ and $X,Y$ amplitudes is given \cite{jemai13}, with references in there. Notably it has been shown how the amplitudes $X, Y$ are solutions of the self-consistent RPA (SCRPA) equations (\ref{RPA-eq}) which we here also call fermion ph-pair mean field equations. Since this is well documented, we will not further comment on this point.\\
We will, however, elaborate a s.p. selfenergy entering the Dyson equation
(\ref{Dyson})
which is consistent
with the SCRPA solution. A self-consistent system of equations can then be established for the calculation of the occupation numbers and the amplitudes $X, Y$, see \cite{Mohsen}. For this, let us consider the dynamic part of the s.p. mass operator given in (\ref{self4}). This contains the following 3-body propagator

\begin{equation}
  G_{k_2k_3k_4;k'_2k'_3k'_4}^{t-t'} = -i \langle 0|{\rm T}(c^{\dag}_{k_2}c_{k_4}c_{k_3})_t(c^{\dag}_{k'_3}c^{\dag}_{k'_4}c_{k'_2})_{t'}|0\rangle
  \end{equation}
We want to establish an integral equation for this propagator. As usual, we employ the equation of motion (EOM) and approximate the integral kernel by the static part. We obtain

\begin{eqnarray}
  (\omega &-&e_{k_3}-e_{k_4} + e_{k_2})G_{k_2k_3k_4;k'_2k'_3k'_4}^{t-t'} = N_{k_2k_3k_4;k'_2k'_3k'_4}\nonumber\\
  &+& \sum K^0_{k_2k_3k_4;l_2l_3l_4}N^{-1}_{l_2l_3l_4;l'_2l'_3l'_4}G_{l'_2l'_3l'_4;k'_2k'_3k'_4}^{t-t'}
  \label{2p1h-prop}
  \end{eqnarray}
where

\begin{equation}
  N_{k_2k_3k_4;k'_2k'_3k'_4} = \langle 0|\{\{c^{\dag}_{k_2}c_{k_4}c_{k_3},c^{\dag}_{k'_3}c^{\dag}_{k'_4}c_{k'_2}\}|0\rangle
  \label{3-body-norm}
\end{equation}
and

\begin{equation}
  K^0_{k_2k_3k_4;k'_2k'_3k'_4} = \langle 0|\{\{c^{\dag}_{k_2}c_{k_4}c_{k_3},[V,c^{\dag}_{k'_3}c^{\dag}_{k'_4}c_{k'_2}]\}|0\rangle
  \label{3b-K}
  \end{equation}
In order to be consistent with our ground state wave function (\ref{Z}), we have to restrict the indices of $K^0$. One namely can show that if one wants to calculate the hole occupation numbers, then the following destruction operator kills $|Z\rangle$ of (\ref{Z}) exactly \cite{o-RPA}

\begin{equation}
  q_{\nu} = \sum_hx^{\nu}_hc^{\dag}_h + \sum_{p_1p_2h}U^{\nu}_{p_1p_2h}c^{\dag}_{p_1}c^{\dag}_{p_2}c_h
  \label{odd-killer}
\end{equation}
where the amplitudes are related by

\begin{equation}
  \sum_h x^{\nu}_hz_{pp'hh'} = U^{\nu}_{pp'h'}; ~~\sum_px^{\nu}_pz_{pp'hh'} = U^{\nu}_{hh'p'}
\end{equation}
and the amplitudes $x, U$ are obtained from the eigenvalue equation corresponding to (\ref{2p1h-prop}) together with (\ref{Dyson})

\begin{equation}
  \begin{pmatrix}{\mathcal  H}_{00}&{\mathcal H}_{01}\\{\mathcal H}_{10}&{\mathcal H}_{11}\end{pmatrix}
  %\begin{pmatrix} x\\U \end{\pmatrix}
  \begin{pmatrix}x\\U \end{pmatrix}
  =E \begin{pmatrix}n_{00}&n_{01}\\n_{10}&n_{11}\end{pmatrix}
  \begin{pmatrix}x\\U \end{pmatrix}
  \label{odd-eigen}
\end{equation}
For the explicit form of the matrix elements, see sect. V.A below.

\begin{figure}
  \resizebox{0.45\textwidth}{!}{\includegraphics{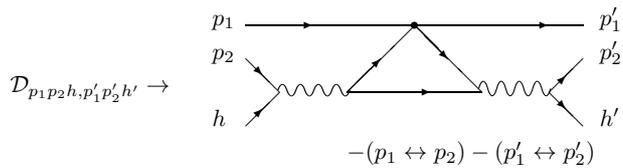}}
  \caption{Schematic view of the PVC interaction vertex which contains itself a selfconsistent PVC process. The full dot stands for the antisymmetrized matrix element and the wiggly line for the vibration. Please note that contrary to the graphical impression, the vertex is instantaneous.}
  \label{PVC-vertex}
  \end{figure}

So the hole destructor couples to a $2p-1h$ configuration. Inversly the particle destructor couples to a 2h-1p configuration. We will only discuss the former, the latter being treated analogously. In the double commutator of (\ref{3b-K}), each of the two triples of 2p-1h fermion operators contracts a particle state to the interaction. Naturally from each triple then only a ph pair remains. We can express those ph pairs via the $Q^{\dag}, Q$ operators of (\ref{Q}) using the inverse relation (\ref{inversion1}). Commuting then the destructors to the right, we exploit the killing property (\ref{killing}) and then $K^0$ is expressed by (self-consistent) RPA amplitudes $X, Y$ and occupation numbers. A graphical representation of this PVC vertex is shown in Fig.\ref{PVC-vertex}.

In turn the occupation numbers are directly related to the s.p. Green's function (11) and then via the dynamical part of the s.p. self energy which is related to the solution of (\ref{2p1h-prop}), we have a closed system of equations for the SCRPA amplitudes, via the SCRPA equations, and the s.p. occupation probabilities. We want to call this sytem of equations the eo-SCRPA. It has been solved for the Lipkin and 1D Hubbard model with very good success as can be seen in the figures presented in sect.V.
From the s.p. Green's function we also can calculate the ground state energy in the usual way

\begin{equation}
  -\frac{i}{2}\lim_{t \rightarrow t' = 0^+}\sum_k[i\frac{\partial}{\partial t} + \tilde e_k]G^{t-t'}_k = E_0
  \label{E0}
  \end{equation}

\begin{figure}
  \resizebox{0.46\textwidth}{!}{\includegraphics{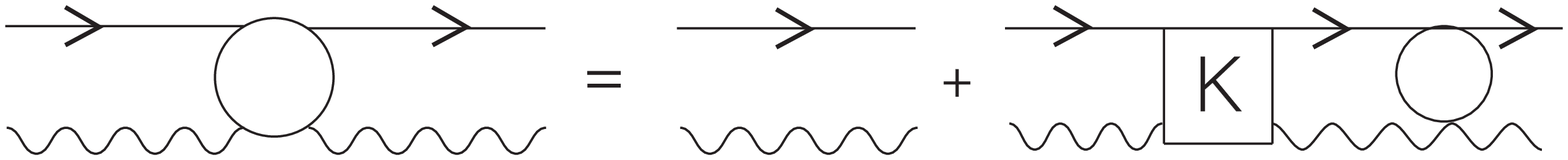}}
  \caption{ Representation of the fermion-vibration scattering equation. The kernel is the one of Fig.\ref{PVC-vertex} which all Pauli exchanges included}
  \label{PVC}
  \end{figure}

Of course, the solution of the $2p-1h (2h-1p)$ equations will in general be quite demanding because of the eventually large configuration space. However, presently in nuclear physics quite routinly in the so-called second RPA huge configuration $2p-2h$ spaces are considered, so that a $2p-1h$ space should be a less important problem. We should also point out that the $2p-1h$ integral equation (\ref{2p1h-prop}) can be interpreted as a particle-vibration scattering eqation with full respect of the Pauli principle. This is schematically shown in Fig.\ref{PVC}. The vibrations (wiggly lines) here are the solutions of the SCRPA equation.

\section{Conservation laws and Ward Identities}

\subsection{Goldstone modes}

It is well established that standard HF-RPA (BCS-QRPA) approach exhibits a Goldstone (zero) mode if the HF solution corresponds to a continuously broken symmetry. For finite systems, one mostly talks about a zero or spurious mode. For nuclei and other selfbound systems like 3He droplets, HF always breaks translational invariance and the corresponding RPA shows a zero mode \cite{RS} which corresponds to a coherent displacement of the whole system. In trapped cold atom gases there exists the so-called Kohn mode where the atom cloud oscillates coherently in an external harmonic trapping potential without internal excitations of the wave packet \cite{Tohy-symm}. Within BCS-QRPA, one obtains in infinite matter, because of the broken particle number $U(1)$ symmetry, a Goldstone mode, the so-called Bogoliubov-Anderson mode \cite{Combescot}. Also in finite superfluid nuclei zero modes appear \cite{Rabhi}.\\

As mentioned, these Goldstone modes reflect basic principles of quantum mechanics and it is very important not to destroy this property in theories which go beyond the HF-RPA scheme. As we see from (\ref{Q}), the crucial point is that the $Q^{\dag}$ operator can represent the symmetry operator (let us call it $\hat S$) in question as a paticular solution of the SCRPA eqauations and that the relation $[H,\hat S] = 0$ is not destroyed  in the course of applying the formalism. In SCRPA {\it all} components of the symmetry operator, besides the diagonal ones, are present and it is important to keep them all \cite{Tohy-symm,Delion-Schuck-Tohy,DSD05,SchuToh,tddm-scrpa}. Therefore, one may think that if the diagonal matrix elements of the symmetry operator $\hat S_{kk}=0 $, then in any case the Goldstone mode will be present in the solution. Many symmetry operators have this property. This is the case for the linear total momentum operator because of its odd parity. In deformed nuclei, the rotational symmetry is broken. The angular momentum operator has no diagonal elements either because it is not time reversal invariant.\\
More subtle is the question of pairing which is one of the broken symmetries often encountered in Fermi systems. In this case the symmetry operator is the particle number operator which contains a Hermitian diagonal piece which cannot be included into the (quasi-particle) RPA operator $Q^{\dag}$ as discussed already. However, in infinite systems the Bogoliubov-Anderson mode comes nevertheless because the diagonal piece of the number operator has zero weight. For finite systems like nuclei one eventually has to extend the theory to cope with the problem, see Ref. \cite{jemai13}.\\
Therefore in practically all situations SCRPA keeps with the Goldstone property. We want to underline the importance of this statement because it is extremely rare to find beyond standard RPA approximations which satisfy the Goldstone theorem besides the multi time (energy) approach for Green's functions discussed by Hedin \cite{Hedin} which, however, is numerically untractable beyond standard RPA.

\subsection{Sum rules}

We show that the energy weighted sum rule, given by the following identity

\begin{equation}
  \sum_{\nu}(E_{\nu} - E_0) |\langle \nu|F|0\rangle|^2 = \frac{1}{2}\langle 0|[F,[H,F]]|0\rangle
  \label{sum-r}
\end{equation}
is fulfilled within SCRPA. Here $|\nu\rangle$ is a complete set of eigenstates
and $F$ is a one body operator

\begin{equation}
  F=\sum_{kk'}f_{kk'}c^{\dag}_kc_{k'}
\end{equation}
One can show that this identity is automatically fulfilled if one considers that $|\nu\rangle$ is the set of SCRPA or r-RPA eigenstates. By using the inverse transformation of the fermion pair operators $c^{\dag}_kc_{k'}$, one obtains

\begin{eqnarray}
  \sum_{\nu}&(&E_{\nu} - E_0) |\langle \nu|F|0\rangle|^2\nonumber\\
  &=&\sum_{\nu}(E_{\nu}-E_0)|\langle 0|[Q_{\nu},F]|0\rangle|^2\nonumber\\
  &=&\sum_{\nu}(E_{\nu}-E_0)|\sum_{kk'}f_{kk'}M^{1/2}_{kk'}(X^{\nu}_{kk'} + Y^{\nu}_{kk'})|^2
\end{eqnarray}
with $M_{kk'} = n_{k'}-n_k$.\\
Using the general system of RPA equations with excitation energies $\Omega_{\nu} = E_{\nu}- E_0$, one can eliminate the amplitudes $X, Y$ in favor of the RPA matrices $A, B$. From there it is only a small step to show that the sum-rule (\ref{sum-r}) is fullilled \cite{Delion-Schuck-Tohy}.

\subsection{Gauge invariance}

Another important property of standard RPA which is fullfilled by SCRPA is gauge invariance (or Ward identity). Gauge invariance of standard RPA is, e.g., nicely demonstrated by Feldman and Fulton \cite{FF}. The extra terms containing the two body correlation functions in (\ref{K-0}) cancel in the limit where the two open legs are put on the same spot in position space. Actually, gauge invariance of standard RPA {\it as well as } SCRPA can easily be verified from (\ref{A,B}), (\ref{Q}).
If in these equations the operator $\delta Q^+_{k_1k_2} = c^{\dag}_{k_1}c_{k_2}/\sqrt{n_{k_2}-n_{k_1}}$ is transformed into $r$-space and the diagonal element is taken, as demanded to verify gauge invariance (see \cite{FF}, eq, (3.69)), we immediately realize that this diagonal operator commutes with the remainder (also written in $r$-space, once the Hamiltonian $H$ is replaced by its interaction part $V$, that is, the Coulomb interaction. Therefore, gauge invariance is fulfilled. This argument is valid discarding spin but, as shown in \cite{FF}, this does not invalidate the general proof. These considerations also entail that the so-called ``velocity-length'' equivalence in the dipole transition is preserved \cite{Delion-Schuck-Tohy}. Please note that all properties mentioned in III.A,B,C are also fullfilled with the renormalized RPA of (\ref{r-ppRPA}), see \cite{Schaefer,SchuToh,tddm-scrpa,Rowe68,Catara}.

\section{Applications}

\subsection{The Lipkin Model}

The Lipkin model, see, e.g., \cite{RS}, is one of the most frequently used models in nuclear physics to test approximation schemes.
\label{Lipkin_Model}
\begin{figure}[ht]
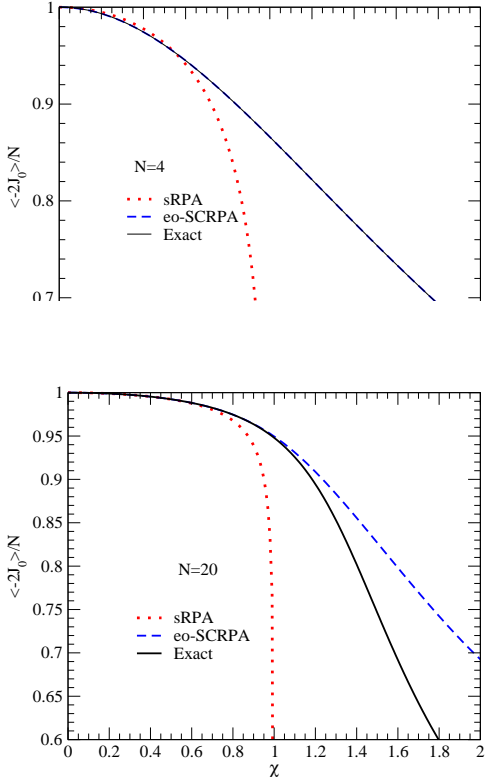

\resizebox{0.35\textwidth}{!}{\includegraphics{newJ0N4.eps}}
\resizebox{0.35\textwidth}{!}{\includegraphics{newJ0N20.eps}}
\caption{\label{newJ0N4N10} The differences between occupation numbers for the various approximation schemes of the two level in Lipkin model, normalized by $N$ as a function of $\chi$ for $N=4,~10,~20,~100$. This with standard RPA (sRPA, dotted red line), eo-SCRPA with eom method for odd particle excitation (broken blue line) and exact solution (full black line). Note that our approach gives the exact result for $N=4$.}
\end{figure}
The single-particle space of the Lipkin model consists of two fermion levels, each of which has a N-fold degeneracy \cite{RS}. The upper (lower) level has the energy of $\frac{e}{2}$ ($-\frac{e}{2}$). The Hamiltonian of the Lipkin model is given by
\begin{equation}
H=e J_0 -\frac{V}{2} \left(J^2_+ +J^2_-\right)
\label{Hlipkin}
\end{equation}
\noindent
with $e $ is the inter-shell spacing, $V$ is the coupling constant, and 
\begin{eqnarray}
J_{0}&=&\frac{1}{2}\sum_{m=1}^{N}\left(c_{1m}^{\dagger}c_{1m}-c_{0m}^{\dagger}c_{0m}\right) , 
\nonumber \\
J_{+}&=&\sum_{m=1}^{N}c_{1m}^{\dagger}c_{0m},, ~~~~~J_{-}=(\hat{J}_{+})^{\dagger } 
\label{su2_operators}
\end{eqnarray}
\noindent
with $2J_0 = \hat{n}_1 -\hat{n}_0 $, $\hat{n}_i =\sum c^{\dagger}_{im} c_{im}$ and $N$ is the number of particles equivalent to the degeneracies of the shells. We consider the odd excitation operator as (\ref{odd-killer}), that is

\begin{equation}
  q^{\dag}_{\mu} = \frac{1}{N}\sum_m x^{\mu}_{0m}c_{0m} + U^{\mu}_{0m}c^{\dag}_{1m}
  \label{q-dag}
\end{equation}
with the  eigenvalue equation of (\ref{odd-eigen}).
Based on the solution of the SCRPA equations \cite{jemai13} with the definition of the ph-pair excitation operator as 
$Q^+= (XJ_+-YJ_-)/d_0$ (with $ d_0=\sqrt{\langle-2J_0\rangle}$ and $J_+= (XQ^+ -YQ )d_0$), we obtain the $X,~Y$ amplitudes as being the solutions of SCRPA equations (\ref{RPA-eq}). The matrix elements of (\ref{odd-eigen}) are then given by

\ba
n_{00}&=&\frac{1}{N}\sum_m\langle\{c_{0m},c^\dagger_{0m}\}\rangle =1
\nn
n_{01}&=&n_{10}=\frac{1}{N}\sum_m\langle\{c_{0m},J_+ c^\dagger_{1m}\}\rangle =0
\nn
n_{11}&=& \frac{1}{N}\sum_m\langle\{c_{1m} J_-,J_+ c^\dagger_{1m}\}\rangle =-\frac{1}{N}(N-2)(1+Y^2)\langle J_0\rangle  
\nn
\H_{00}&=&\frac{1}{N}\sum_m\langle\{c_{0m},[H,c^\dagger_{0m}]\}\rangle = -\frac{e}{2}
\nn
\H_{10}&=&\H_{01}=\frac{1}{N}\sum_m\langle\{ c_{1m}J_-,[H,c^\dagger_{0m} ]\}\rangle =-V n_{11}
\nn
\H_{11}&=&\frac{1}{N}\sum_m \langle\{ c_{1m}J_-,[H,J_+c^\dagger_{1m} ]\}\rangle
\nn
&=& \frac{3e}{2}n_{11} - 2VXY(2-\frac{8}{N})[(1+2Y^2)\langle J_0\rangle +\langle J_0^2\rangle]
\ea
and the corresponding secular equation 
%{\bf check si l'ecriture de l'eq est correcte} Ok
\begin{eqnarray}
\det\left(\sum_{i'j'}\N^{-1/2}_{ii'}\H_{i'j'}\N^{-1/2}_{j'j}-\lambda I\right)= 0 
\label{matrixeqlipk}
\end{eqnarray}
The occupations numbers are given by

\begin{eqnarray}
n_{0} &=& N \frac{\lambda _- -\H_{11}/n_{11}}{\lambda_{-} -\lambda _+}
~~~\mbox{and}~~~
n_1 = N-n_0~~~
\end{eqnarray}
where $\lambda_{\pm}$ are the eigenvalues of the $2\times 2$ matrix problem,
\begin{eqnarray}
\lambda _\pm &=& -\frac{e}{2} +\beta\pm \sqrt{\beta ^2+V^2n_{11}}
\label{valpropLipk}
\end{eqnarray}
\noindent
with $\beta =e- VXY(N-4)-VXY(N-4)(1+2Y^2)\frac{\langle J_0\rangle}{n_{11}}$. Thus,
\begin{eqnarray}
\langle -2J_0\rangle = n_0 -n_1 = 2n_0 -N
\label{2j0Wx0}
\end{eqnarray}

\begin{figure}[ht]
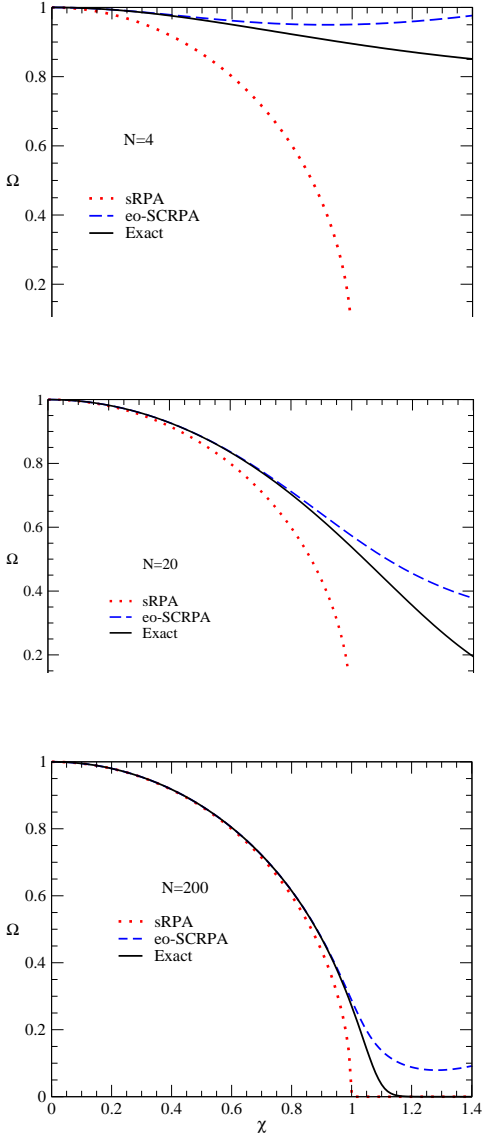

\resizebox{0.35\textwidth}{!}{\includegraphics{new_AnsatzN4.eps}}
\resizebox{0.35\textwidth}{!}{\includegraphics{new_AnsatzN20.eps}}
\resizebox{0.35\textwidth}{!}{\includegraphics{new_AnsatzN200.eps}}
  \caption{\label{omegaN8_density} 
The first excited state of even systems with different approximations as sRPA and eo-SCRPA compared to exact solution as a function of $\chi$ for $N=4,~20,~200$. Please note that one may make the hypothesis that the eo-SCRPA approach becomes exact in the $N\rightarrow \infty$ limit.}
\end{figure}

\begin{figure}[ht]
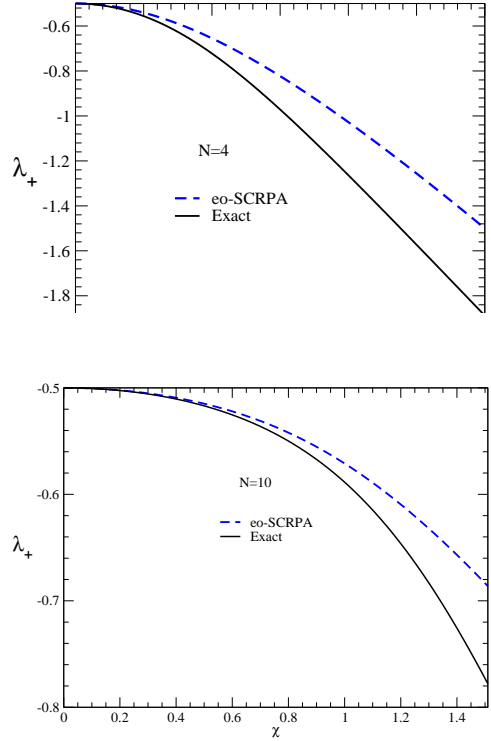

\resizebox{0.35\textwidth}{!}{\includegraphics{spectN4P1oSCRPA.eps}}
\resizebox{0.35\textwidth}{!}{\includegraphics{spectN10P1oSCRPA.eps}}
  \caption{\label{SPECTN4P1oSCRPA} Excitation energy between the system $N+1$ and $N$ particles as a function of $\chi =V(N-1)$ for $N=4, ~10$ with eo-SCRPA (blue dashed line) (\ref{valpropLipk}) compared to the exact solution (full black line) $\lambda_{+}= E^{N+1}_{\alpha}-E^{N}_0$.  }
\end{figure}
\begin{figure}[ht]
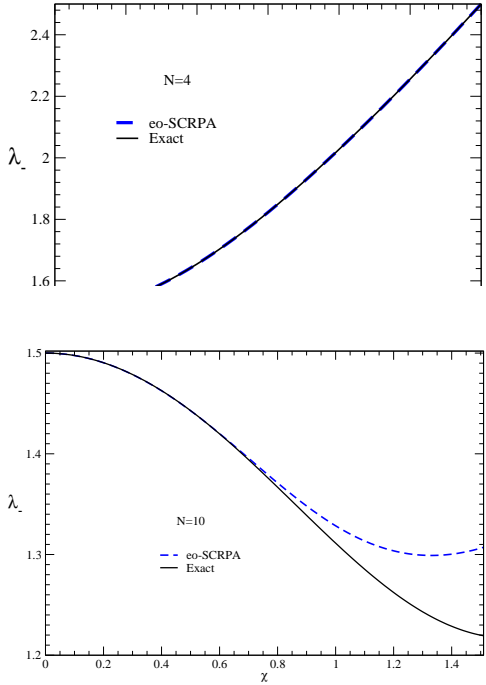

\resizebox{0.35\textwidth}{!}{\includegraphics{spectN4M1oSCRPA.eps}}
\resizebox{0.35\textwidth}{!}{\includegraphics{spectN10M1oSCRPA.eps}}
  \caption{\label{SPECTN4M1oSCRPA} Same as Fig. \ref{SPECTN4P1oSCRPA} but for the excitation energy between the system $N-1$ and $N$ particles as a function of $\chi=V(N-1)$ for $N=4, ~10$. Note that for $N=4$ the exact result $\lambda_{-} =E^{N- 1}_{\alpha}-E^{N}_0$ is obtained with our approach eq. (\ref{valpropLipk}).  }
\end{figure}
\noindent
In the above equations (\ref{matrixeqlipk}) the correlation functions are expressed by the RPA amplitudes $X, Y$ in the way it is described in section III. The correlation functions which contain quadratic forms of occupation number operators as $\langle J_0J_0\rangle$ in above equation can in principle be expressed by the RPA amplitudes as well but leading to heavier expressions. Usually, we, therefore will employ the factorisation approximation leading in the present case to $\langle J_0J_0\rangle \simeq \langle J_0\rangle^2$ what mostly turns out to be quite satisfactory. However, in the case of the Lipkin model one also can use the Casimir relation

\[ 4\langle J_0^2\rangle = N(N+2)+4\langle J_0\rangle -4\langle J_+J_-\rangle \]
to close the systme of equations. The results are shown in Figs.\ref{newJ0N4N10} and in \cite{Mohsen}. They concern in the order: i) the expectation value $\langle J_0\rangle$ of the difference of populations in upper and lower level, ii) the first excitation energy of the even system, iii) the excitation energies of the odd system. The correlation energy (not shown) is reproduced with  the same quality. All quantities are very well reproduced throughout couplings up to the  critical value $\chi = \chi_{\rm crit.}$ where the standard RPA breaks down and the system wants to change to the 'deformed' basis. However, even values slightly beyond $\chi_{\rm crit.} =1$  are still quite acceptable. All quantities for $N=2$ are reproduced exactly. By some lucky accident the occupancies even for $N=4$ come out to be exact (as shown in Fig.\ref{newJ0N4N10}, Fig.\ref{SPECTN4M1oSCRPA} and in \cite{Mohsen}).

\subsection{The Hubbard Model}
\label{HubbMod}

The Hubbard model is widely used to deal with the physics of strongly correlated electrons. Since the model can be solved exactly in one dimension (1D) and for small
cluster sizes, it is very useful for theoretical investigations \cite{jemai05,Hubbard}. To be precise, our "Hubbard model" is a 6-site system at half filling with periodic boundary condition, described by the usual Hamiltonian \cite{jemai05,Hubbard}: 
\begin{figure}[ht]
\resizebox{0.25\textwidth}{!}{\includegraphics{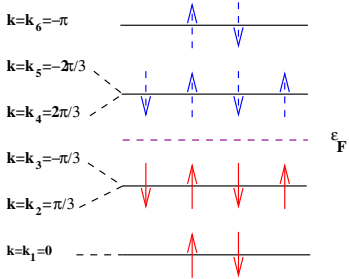}}
  \caption{\label{spectrumHF6sites} Hatree Fock States at $U=0$ for the chain with 6 sites at half filling and projection of spin $m_{s} = 0$. The occupied states are represented by the full arrows and those not occupied are represented by the dashed arrows. }
\end{figure}
\begin{equation}
H=-t\sum\limits_{\langle i,j\rangle ,\sigma }c_{i\sigma }^{\dagger
}c_{j\sigma }+\frac{U}{2}\sum\limits_{i,\sigma} \hat{n}_{i,\sigma }\hat{n}_{i,-\sigma }.
\label{Hamiltonian_hubbard}
\end{equation}
Here, $\hat{n}_{i\sigma }=c_{i\sigma }^{\dagger }c_{i\sigma }$, $c_{i\sigma }^{\dagger }$ and $c_{i\sigma }$ are the creation and annihilation operators for an electron at site $i$ with spin $\sigma $, $U$
is the on-site (spin-independent) interaction, $-t$ is the hopping term of the kinetic energy. The eigenstates of the system will be linear combinations of Slater determinants, which are denoted by the kets $|1\ldots 6\rangle $, with occupations of the sites $1\ldots 6$.
The hamiltonian is rewritten in plane wave basis, 
\ba
H&=&\sum_{{\bf{k}} \sigma} \varepsilon_{\bf{k}} \hat{n}_{{\bf{k}} \sigma}
+\frac U{2N}\sum_{\bf{kk'q}\sigma} a^{\dagger}_{{\bf{k}}\sigma} a_{{\bf{k+q}}\sigma}
a^{\dagger}_{{\bf{k'}} -\sigma} a_{{\bf{k'-q}} -\sigma} ~~~~~
\label{ham_imp}
\ea
with the transformation
\be
c_{j,\sigma}=\frac 1{\sqrt{N}} \sum_{\bf{k}} a_{\bf{k},\sigma} e^{-i\bf{k\,x_{j}}} ~\mbox{,} \label{6sit_trans_conssym0ca}
\ee
where $\hat{n}_{\bf{k}, \sigma} = a^{\dagger}_{\bf{k}, \sigma} a_{\bf{k}, \sigma}$, $\varepsilon _{\bf{k}}= -2t \cos\left(ka\right)$, which are, respectively, the number operator of particles of the mode $({\bf{k}},\,\sigma)$ and the energies of one particle on a lattice with a the parameter of the lattice which is taken as $a=1$. For a problem with $N$sites, the condition of periodicity is given by $c_{N+1,\sigma}=c_{1,\sigma}$. This implies that $e^{-ik\,N}=1$, hence the values taken by $k$ will be $k=\frac {2\,\pi}{N}\,n$. In addition, the first Brillouin zone is defined on the field where $-\pi\leq k <\pi$, which gives us the values of $n$ as $\frac{-N}{2}\leq n <\frac{N}{2} $.

For the six sites, we have the possible states with the following wave vectors:
\ba
k_1=0,~k_3=-k_2=\frac{\pi}{3} ,~k_5=-k_4=\frac{2\pi}{3} ,~k_6=-\pi
\ea 
and with the kinetic energies (see Fig.\ref{spectrumHF6sites}), respectively,
\ba
\varepsilon_{k_{6}}=-\varepsilon_{k_{1}}=2\,t,~~
\varepsilon_{k_{4}}=\varepsilon_{k_{5}}=-\varepsilon_{k_{2}}=-\varepsilon_{k_{3}}=t .
\ea 
\noindent
The transfer wave vector($q_{ph}=k_p-k_h$) takes the possible values as shown in the Table \ref{table1}.
\begin{table}[h]
\begin{tabular}{|c|c|c|}
\hline
$q =\pm\frac{2\pi}{3}$  & $q =\pm \pi$ & $q =\pm \frac{\pi}{3}$
\\ \hline
$51\rightarrow q_{51} =+\frac{2\pi}{3}$ & $61\rightarrow q_{61}  =-\pi$ & $42\rightarrow q_{42} =-\frac{\pi}{3}$
\\ \hline
$63\rightarrow q_{63} =+\frac{2\pi}{3}$ & $52\rightarrow q_{52}  =+\pi$ & $53\rightarrow q_{53} =+\frac{\pi}{3}$
\\ \hline
$41\rightarrow q_{41} =-\frac{2\pi}{3}$ & $43\rightarrow q_{43}  = -\pi$ &
\\ \hline
$62\rightarrow q_{62} =-\frac{2\pi}{3}$ &  &
\\ \hline
\end{tabular}
\caption{\label{table1} The various momentum transfers in the 6 sites case.}
\end{table}
\begin{figure}[ht]
\resizebox{0.35\textwidth}{!}{\includegraphics{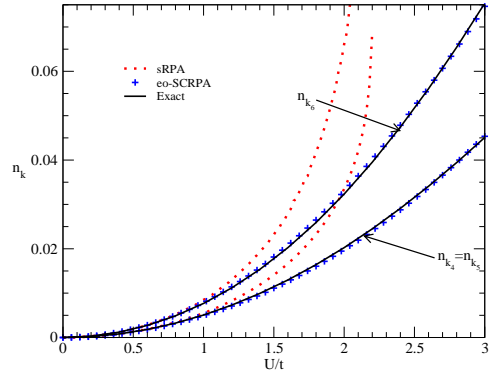}}
  \caption{\label{Hcomplet} Occupation numbers as function of the interaction $U/t$ for various values of the momenta $k_6=-\pi$, $k_5=-2\pi/3$, $k_4=2\pi/3$ for states above the Fermi level. Notice that the modes $k_4 = 2\pi/3$ and $k_5 =-2\pi/3$ are degenerate. For each approximation, sRPA (red dots) and eo-SCRPA (blue crosses), are  compared to the exact solution (full black line).  Also we have $n_{k_1}=1-n_{k_6}$ and $n_{k_2}=n_{k_3}=1-n_{k_4}=1-n_{k_5}$ }
\end{figure}
\begin{figure}[ht]
\resizebox{0.35\textwidth}{!}{\includegraphics{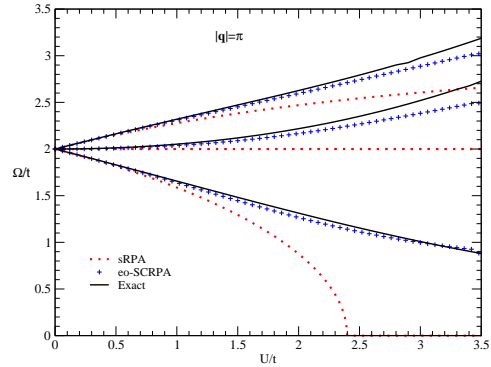}}
\caption{\label{rep1p1} Same as Fig.\ref{Hcomplet} but for the energy of the excited state in the channel $|q| = \pi $. The channels $|q| =\pi/3$, $2\pi/3 $ are of equal quality.}
\end{figure}

At this point we proceed exactly as in the case of the Lipkin model: The excitation operator for the even system is given by
\ba
Q^{\dagger}_{\nu} = \sum_{ph\sigma}\X^{\nu}_{ph\sigma}\, K^{+}_{ph\sigma}-\Y^{\nu}_{ph\sigma}\, K^{-}_{hp\sigma}
\label{op-dexqp6s4}
\ea
with $~ K^{\pm}_{ph\sigma }=J^{\pm}_{ph\sigma}/\sqrt{N_{ph\sigma}} $, $~J^{+}_{ph\sigma}=a^{\dag}_{p\sigma}a_{h\sigma}$, $~N_{ph\sigma}=n_{h\sigma}- n_{p\sigma}$. With the inversion
\be
%J^{-}_{hp\sigma} & = & \sqrt{N_{ph\sigma}}\;\sum\limits_{\nu}\;\left(\;\X^{\nu}_{ph\sigma} \; Q_{\nu} + \Y^{\nu}_{ph\sigma} \; Q_{\nu}^{\dagger}\;\right)
%\nonumber \\
J^{+}_{ph\sigma} = \sqrt{N_{ph\sigma}}\sum\limits_{\nu}\;\left(\;\Y^{\nu}_{ph\sigma} \; Q_{\nu} + \X^{\nu}_{ph\sigma} \; Q_{\nu}^{\dagger}\;\right)~.
\label{inversion_1}
\ee
we can calculate the mean values needed for the matrix elements of the SCRPA equations for the even particle number case
\begin{equation}
  \langle J^{+}_{p'h'\sigma '}\,J^{-}_{hp\sigma}\rangle =
  \sqrt{N_{p'h'\sigma '} N_{ph\sigma}}\; \sum\limits_{\nu}\;\Y^{\nu}_{p'h'\sigma '}\; \Y^{\nu}_{ph\sigma},
\end{equation}
and similar expressions for other expectation values of this type
%\mbox{,}
%\nonumber \\
%\langle J^{-}_{h'p'\sigma '}\,J^{+}_{ph\sigma}\rangle & = &
%\sqrt{N_{p'h'\sigma '} N_{ph\sigma}}\; \sum\limits_{\nu}\;\X^{\nu}_{p'h'\sigma '}\; \X^{\nu}_{ph\sigma}
%\mbox{,}
%\nonumber \\
%\langle J^{+}_{p'h'\sigma '}\,J^{+}_{ph\sigma}\rangle & = &
%\sqrt{N_{p'h'\sigma '} N_{ph\sigma}}\; \sum\limits_{\nu}\;\Y^{\nu}_{p'h'\sigma '}\; \X^{\nu}_{ph\sigma}
%\mbox{,}
%\nonumber \\
%\langle J^{-}_{h'p'\sigma '}\,J^{-}_{hp\sigma}\rangle & = &
%\sqrt{N_{p'h'\sigma '} N_{ph\sigma}}\; \sum\limits_{\nu}\;\X^{\nu}_{p'h'\sigma '}\; \Y^{\nu}_{ph\sigma}
%\mbox{,}~~~\label{foctcorr_jj}
%\ea
where we replaced the "ph" operators by the RPA creation and destruction operators from the inversion (\ref{inversion1}) and then commute the $Q$ operators to the right until they kill the ground state. All matrices become functionals of the occupancies $n_h$ and $n_p$ and $X, Y$ amplitudes in analogy to what was the case in the Lipkin model and, thus, the diagonalisation process implies at the same time an iteration on the occupancies and the amplitudes.\\
For the odd particle number case, we make again the following ansatz
\begin{eqnarray}
q^\dag_{h,\mu} &=& x^\mu_{h} a_{h+} +\sum_{p'ph} U^\mu_{p'ph} a^\dagger_{p'+} J^+_{ph-}
\nn
q^\dag_{p,\rho}&=& x^\rho_{p} a^\dagger_{p+} + \sum_{p'h'h} U^\rho_{p'h'h} a^\dagger_{h+}J^-_{h'p'-}  ~.
\end{eqnarray}
From there, we can, as outlined in the general section II, and as just now for the case of the Lipkin model, calculate the occupation numbers. The results for the occupation numbers are again excellent, see Fig.~\ref{Hcomplet}. Also the excitation energies of the even particle number system, see Fig.~\ref{rep1p1} are very well reproduced.\\
%In Fig.~\ref{EGS_Hubb6ph} we show the ground state energies for the exact case compared to the eo-SCRPA solution.\\

\section{Conclusion}

In this work, we coupled even and odd particle numbers RPA selfconsistently. Both systems are based on the same correlated RPA ground state. From the odd system, we get the occupation numbers, odd particle excitation energies, and the ground state energies whereas from the even SCRPA equations we get the excitation energies of the even system and transition probablities. Both even and odd systems are coupled through non-linear equations. We called this system of equations 'even-odd' SCRPA (eo-SCRPA). Applications to the Lipkin model and a six sites Hubbard ring at half filling gave excellent results for all quantities. The equations are relatively complex due to their non-linearity but they should be solvable with modern computers for realistic problems. The coupling of even and odd RPA's has a couple of advantages: it gives richer results, i.e., excitation energies of even and odd particle number systems; there is a natural way how to obtain the ground state energy via the s.p. Green's function and, last but not least, all qualities of standard RPA, as there are, the Goldstone theorem, sum-rules, and gauge invariance, respectively Ward identities are maintained. Since there is no phenomenological input in the eo-SCRPA equations and the hard core is tamed by the $T$-matrix, our scheme can be qualified as an ``ab initio'' PVC approach. The results of the Lipkin and Hubbard models seem to be very promising. Work for realistic applictions is planned for the future \cite{Litvinova}.
\section{Acknowledgements}
I am very greatful to M. Jemai and Zhou Bo for their help with the preparation of the figures and M. Jemai for collaboration on this subject.
I appreciate longstanding collaboration on SCRPA with D. Delion, J. Dukelsky, and M. Tohyama. Thanks are also due to G. R\"opke for discussions concerning the pp-channel. V. Olevano and J. Toulouse are collaborators of a very fruitful recent work and publication in the ph-channel \cite{Julien}.

\appendix

\section{Equation of Motion for odd particle number operator for Hubbard Model}
\label{AppHubbMod}
\noindent
For the Hubbard model (\ref{ham_imp}) we define the odd excitation operator as in Eq.(\ref{odd-killer}),
\begin{eqnarray}
q^\dag_{h,\mu} &=& x^\mu_{h} a_{h+} +\sum_{p'ph} U^\mu_{p'ph} a^\dagger_{p'+} J^+_{ph-}
\nn
\nn
q^\dag_{p,\rho}&=& x^\rho_{p} a^\dagger_{p+} + \sum_{p'h'h} U^\rho_{p'h'h} a^\dagger_{h+}J^-_{h'p'-}  ~.
\end{eqnarray}
\noindent 
with $J^+_{ph-} =a^\dagger_{p-} a_{h-}$ and $\sigma =\uparrow, \downarrow=+,-$.
Remembering the  notations for the occupation probabilities
\begin{eqnarray}
n_{k\sigma}=\langle\hat{n}_{k\sigma}\rangle &=& \langle a^\dagger_{k\sigma} a_{k\sigma}\rangle , 
%~,~~~~~~~\tilde{n}_k = \langle b^\dagger_{k} b_{k}\rangle ~,~~~~~~~
%\hat{n}_p =\tilde{n}_p ~,~~~~~~~ \hat{n}_ h = 1- \tilde{n}_h
\end{eqnarray}
we have $n_{k_2\sigma} =n_{k_3\sigma}$, $n_{k_4\sigma} =n_{k_5\sigma}$, $n_{k_2\sigma} =1-n_{k_3\sigma}$ and $n_{k_1\sigma} =1-n_{k_6\sigma}$. This gives 
\begin{eqnarray}
\H_{11}=\langle\{a_{k_1+},[H, a^\dagger_{k_1+}]\}\rangle = \epsilon_{k_1} =-2t +U/2
\end{eqnarray}
\noindent
The term without interaction $H_{0}=\sum_{k\sigma} \varepsilon_{k} \hat{n}_{k\sigma} $ is given by 
\begin{eqnarray}
\langle \{ a_{p'+} J^-_{hp-},[H_{0}, a^\dagger_{p'+} J^+_{ph-}] \} \rangle 
&=&(\varepsilon_{p} -\varepsilon_{h} +\varepsilon_{p'})\N_{p'ph} \nn
\ea
with $\N_{p'ph}=\langle (1-\hat{n}_{p'+})(-2 J^0_{ph,-})\rangle +\langle J^+_{ph,-} J^{-}_{hp,-} \rangle$. The term in the Hamiltonian for the transfer $q=0$, $H_{q=0}=\frac U{6}\sum_{kk'} \hat{n}_{k+} \hat{n}_{k' -} $ leads to
\ba
\langle \{ a_{p'+} J^-_{hp-},[H_{q=0}, a^\dagger_{p'+} J^+_{ph-}] \} \rangle  &=& \frac{U}{2}\N_{p'ph} 
\end{eqnarray}
with $\sum_k \hat{n}_{k\sigma} =\sum_p \hat{n}_{p\sigma} +\sum_h \hat{n}_{h\sigma} =3 $ in the half-filled case. Now let us calculate the elements $\C_{p'ph}$ for the first row (or column) as
\ba
\sqrt{\N_{p'ph}}\C^*_{p'ph,h_1} &=&\langle\left\{a_{p'+} J^-_{hp-}, \left[H,a^\dagger_{h_1+} \right]\right\}\rangle
\nn
&=&\frac{U}{6}\biggr\{\langle a^\dagger_{h_1-q+}a_{p'+} a^\dagger_{h+q-}a_{p-}\rangle
\nn
&&
-\langle a^\dagger_{h_1-q+}a_{p'+} a^\dagger_{h-}a_{p-q-}\rangle
\nn
&&
+\sum_{k} \langle J^-_{hp-} a^\dagger_{k-} a_{k+p'-h_1-}\rangle
 \biggl\}
\ea
The elements of the matrix except the first row (or column) are given as follows 
\onecolumn
\ba
\sqrt{\N_{p'ph}\N_{p''p_1h_1}}\D_{p'ph,p''p_1h_1} &=& \langle \{ a_{p''+} J^-_{h_1p_1-},[H, a^\dagger_{p'+} J^+_{ph-}] \} \rangle
\nn
&=&(\epsilon_{p}+\epsilon_{p'}-\epsilon_{h})\dl_{p'p''}\biggl\{\langle J^-_{h_1p_1-}  J^+_{ph-}\rangle  +\dl_{hh_1}\dl_{pp_1}\langle \hat{n}_{p'+}(\hat{n}_{p-}-\hat{n}_{h-})\rangle \biggr\}
\nn
&&+\frac{U}{6}\dl_{p'p''}\biggl\{\sum_{kq}
 \langle a^\dagger_{k+}a_{k+q+} J^-_{h_1p_1-} (a^\dagger_{p+q-} a_{h-}-a^\dagger_{p-} a_{h-q-})\rangle \biggr\}
\nn
&&+\frac{U}{6}\dl_{p'p''}\dl_{hh_1}\biggl\{\sum_{kq}
 \langle a^\dagger_{k+}a_{k+q+} \hat{n}_{p'+} a^\dagger_{p+q-}a_{p_1-}\rangle 
-\sum_{k}\langle a^\dagger_{k+}a_{k+p_1-p+} \hat{n}_{p'+} \hat{n}_{h-}\rangle \biggr\}
\nn
&&+\frac{U}{6}\dl_{p'p''} \dl_{pp_1}\biggl\{\sum_{kq}
\langle a^\dagger_{k+}a_{k+q+} \hat{n}_{p'+} a^\dagger_{h_1-}a_{h-q-}\rangle 
-\sum_{k}\langle a^\dagger_{k+}a_{k+h-h_1+} \hat{n}_{p'+} \hat{n}_{p-}\rangle \biggr\}
\nn
&&+\frac{U}{6}\dl_{pp_1}\dl_{hh_1}\biggl\{\sum_{kq}
\langle a^\dagger_{p'-q+}a_{p''+}\hat{n}_{p-}a^\dagger_{k-}a_{k-q-}\rangle
-\sum_{kq}\langle a^\dagger_{p'-q+}a_{p''+}\hat{n}_{h-}a^\dagger_{k-}a_{k-q-}\rangle 
\biggr\}
\nn
&&+\frac{U}{6}\biggl\{\sum_{q}
\langle a_{p''+q+}a^\dagger_{p'+} J^-_{h_1p_1-} (a^\dagger_{p+q-} a_{h-}- a^\dagger_{p-} a_{h-q-})\rangle
\nn
&&~~~~~+\sum_{q}\langle a^\dagger_{p'-q+}a_{p''+} J^+_{ph-} (a^\dagger_{h_1+q-}a_{p_1-}-a^\dagger_{h_1-}a_{p_1-q-})\rangle 
\nn
&&~~~~~+\sum_{k}\langle J^-_{h_1p_1-} J^+_{ph-} a^\dagger_{k-} a_{k-p'+p''-}\rangle\biggr\}
\label{triplet}
\ea

\noindent
In the following, as already discussed several times, we retain from (\ref{triplet}) only those terms where the particle states of the left and right triple operators in $\D$ connect to the interaction. The remaining density operator from the interaction is approximated by its diagonal form. This leads to expressions evaluated in (\ref{odd-expectation}) below. First let us discuss what kind of terms we are neglecting in this way.
It should be noted that the terms of type $\langle J^{\pm}_{ph}J^{\pm}_{p'h'}J^{\pm}_{p''h''}\rangle =0$, $\langle J^{\pm}_{ph}S_{p_1p_2}J^{\pm}_{p''h''}\rangle$ are probably small (with $S_{p_1p_2}=a^\dagger_{p_1} a_{p_2}$ for $p_1\neq p_2$) and $\langle J^{\pm}_{ph}S_{h_1h_2}J^{\pm}_{p''h''}\rangle$ also small (with $S_{h_1h_2}=a^\dagger_{h_1} a_{h_2}$ for $h_1\neq h_2$)  in eq.(\ref{triplet}). As shown in \cite{jemai13}, the term $\langle SJ\rangle =0$ and $\langle SS\rangle$  are small. Only the terms nonzero in eq.(\ref{triplet}) like $\langle J^{\pm}_{ph+}n_{k\pm} J^{\pm}_{p'h'-}\rangle $ which can be calculated as shown  in (\ref{odd-expectation}).
%Then, We can calculate the terms non-zero in eq.(\ref{triplet}). 
With the short hand notation $ph\sigma \equiv i$, $k\sigma \equiv k$, $\hat{N}_i=\hat{n}_{h\sigma} -\hat{n}_{p\sigma}$ and $N_i=n_{h\sigma} -n_{p\sigma}$, we can evaluate the following terms

\ba
\langle J^-_{i}\hat{n}_{k} J^-_{j}\rangle &=&\sqrt{N_iN_j} \sum_{\nu,\nu '} X^\nu_i Y^{\nu '}_j \langle Q_\nu \hat{n}_{k} Q^\dag_{\nu '} \rangle
\nn
&=&\sqrt{N_iN_j} \sum_{\nu,\nu '} X^\nu_i Y^{\nu '}_j \left( X^\nu_i X^{\nu '}_j  -Y^\nu_i Y^{\nu '}_j \right) 
+ \sum_{\nu} X^\nu_i Y^{\nu }_j \sum_{l} \left( |X^\nu_l|^2 -|Y^\nu_l|^2\right) \langle \hat{n}_{k} \hat{N}_l \rangle
\label{odd-expectation}
\ea
and other mean values of similar type.

\bibliographystyle{refer}
\bibliography{articles}

\end{document}